\documentclass[conference]{IEEEtran}
\IEEEoverridecommandlockouts
\usepackage{balance}
\usepackage{graphicx}
\usepackage{multicol}

\PassOptionsToPackage{hyphens}{url}\usepackage{hyperref}
\usepackage{url}
\usepackage{enumitem}
\usepackage{graphicx}

\makeatletter
\IEEEtriggercmd{\reset@font\normalfont\scriptsize}
\makeatother
\IEEEtriggeratref{1}

\usepackage[a4paper]{geometry}
\begin{document}

\title{Smart Humans... WannaDie?}

\author{\IEEEauthorblockN{Diego Sempreboni}
\IEEEauthorblockA{\textit{Department of Informatics}\\ 
\textit{King's College London}, London, UK \\
diego.sempreboni@kcl.ac.uk}
\and
\IEEEauthorblockN{Luca Vigan\`{o}}
\IEEEauthorblockA{\textit{Department of Informatics}\\ 
\textit{King's College London}, London, UK \\
luca.vigano@kcl.ac.uk}
}

\maketitle

\begin{abstract}

It won't be long until our prostheses, ECG personal monitors, subcutaneous insulin infusors, glasses, etc. become devices of the Internet of Things (IoT), always connected for monitoring, maintenance, charging and tracking. This will be the dawn of the Smart Human, not just a user of the IoT but a Thing in the Internet. 
How long would it then take for hackers to attack us like they have been attacking IoT devices?
What would happen if hackers were able to blackmail us threatening our IoT body parts? Smart Humans may become victims of the devastating attack of \emph{WannaDie}, a new ransomware that could provide the plot-line for a possible future episode of the Black Mirror TV series.
\end{abstract}

\section{The Internet of Things}

The \emph{Internet of Things (IoT)} has proved to be one of the biggest technological and societal revolutions since the ``opening'' of the ARPANET project~\cite{ARPANET}. 
IoT is a network of physical devices, vehicles, home appliances and other items embedded with electronics, software, sensors, actuators, and connectivity that enables these objects to connect and exchange 
data~\cite{Al-Fuqaha2015,mandula2015mobile,osisanwo2015internet}. 
It is estimated that by 2020 there will be 30 billion devices, with the global market value of IoT reaching \$7.1 
trillion~\cite{IoT2020}. 

IoT has been creating opportunities for more direct integration of the physical world into computer-based systems, resulting in efficiency improvements, economic benefits, and reduced human exertions~\cite{LePallec}.
Nowadays, almost each everyday object can become a ``Thing'' of the IoT when it is connected to the net. However, by providing such possibilities IoT has also been unearthing a huge number of security and privacy problems. 

Although it can be questionable whether connecting a kettle, a toaster or a toothbrush to the net is actually really useful, most IoT applications have been welcomed by the community of users (and developers).
Think, for instance, of the introduction of smart meters in our homes along with controllable devices such as thermostats, lights and shutters. Or smart TVs and fridges, which have been changing, and improving, the way in which we access entertainment and shop for food in a sustainable way.  The benefits of IoT are even more substantial and evident for industry and infrastructures: energy and smart grid, manufacturing, food supply chain, transport and logistics are some of the areas that have been impacted by IoT~\cite{da2014internet}. For instance, the courier company DHL highlighted the following advantages of using the IoT in logistics~\cite{macaulay2015internet}: vehicle monitoring and maintenance, environmental sensors in shipping containers, real-time tracking of packages, information-gathering on employees and tools, and a number of safety-enhancing features for vehicles and people. Cisco, a leader in IT and networking, have also been advocating ``Industrial IoT''~\cite{cisco}: remote monitoring and access to the equipment used in manufacturing could greatly improve efficiency and allow issues to be resolved more quickly, thereby increasing production.

There are also plenty of applications of the IoT in healthcare; for instance, RFID technology has been applied to passive (i.e., battery-less) devices for monitoring a patient's local environment. Pacemakers too have become IoT devices: 
it is nowadays possible to adjust a pacemaker's configuration through an external control device, without invasive surgery. 

\section{Enhancing Humans Through Technology}

Humans have always tried to improve themselves in different ways. By studying and learning, by training their bodies, but also by ``enhancement''. We may distinguish two concrete ways that humans have pursued to enhance themselves: the \emph{biological way} and the 
\emph{technological way}.

For what concerns the biological way, as Darwin and others taught us, humans, like other species, have been undergoing a continuous and likely never-ending evolutionary process. 
Changes to the habitat and climate, as well as changes driven by specific needs, have forced humans to adapt 
and evolve. Life expectancy has been drastically increased thanks to breakthroughs in science and medicine, such as advances
in treatment and prevention of diseases, ground-breaking organ transplantation techniques and general progress in health-care. 

Complementary to the biological way, we have nowadays the technological way.
Thanks to various breakthroughs, many of which were utterly unimaginable until recently, technology has given us the ability to alter biology, along with the social conditions and cultural expectations that enable such transformations. 
There are a variety of enhancement technologies that augment or improve bodily shapes and functions with the aim of improving human characteristics, abilities and skills, including appearance and mental or physical functioning. Although some of these technologies may sometimes have been applied beyond what is ``normal" or necessary for well-being~\cite{Extremebiohacking}, in most cases they have made a huge difference. 

For instance, in addition to new medical and pharmacological discoveries, technology has been used to treat, monitor and relieve certain diseases. Type 1 diabetes is being treated using an artificial pancreas device system~\cite{fda}, where a system of devices (a monitor, an insulin infusion pump and a blood glucose meter)
closely mimic the glucose-regulating function of a healthy pancreas. Low-invasive and highly technological pacemakers are being used for the treatment of arrhythmias and dysfunctions of the heart. Robotic prosthetic arms and legs are being used to restore 
abilities to patients who were born without limbs or lost them in accidents~\cite{darpa}.
Some patients affected by paralysis can now walk again thanks to an implanted device~\cite{vergewalk} or thanks to exoskeletons, which are being used also for military and civil applications~\cite{bogue2009exoskeletons,pratt2004roboknee}. 

\section{Wait a minute! What about security?}

Like for many technological advancements, IoT is an application-driven field in which innovation is being pushed forward for the large part by non-tech people rather than by real and significant needs. Consumers want smart microwaves~\cite{alexamicrowaves}, smart lights, smart lockers, smart toasters and so on. Basically, the formula is the following one: take an existing product, add \textit{smart} in front of its name by allowing it to be connected to the Internet, produce it, sell it, cash in. It is not up to us to judge whether this race towards ``smartification'' is sensible or not. As we observed, there are some remarkable advantages, and some serious doubts whether things have been taken too far (or whether the Internet of Things has been taken so far).
But, most importantly, there are major security problems. We have once again lost the opportunity to consider security from the start like what happened when Internet was first designed\footnote{As pointed out Danny Hills in~\cite{LoAndBehold}: ``Because the Internet was designed for a community that trusted each other, it didn't have a lot of protections in it. We didn't worry about spying on each other, for example. We didn't worry about somebody sending out spam, or bad emails, or viruses, because such a person would have been banned from the community.''} and has happened for every major ``update'' of the Internet.

As history teaches us, the frantic rush to be the first on the market may mean little or even no security at all. Vendors are hastily seeking to dish out the next innovative connected gadget before competitors do. Under such circumstances, functionality becomes the main focus and security takes a back seat. Indeed, new attacks on IoT devices are being reported almost on a daily basis. Philips Hue smart lightbulbs were one of the first IoT devices on the market, and one of the first to be attacked~\cite{dhanjani2013security}, even remotely~\cite{ronen2017iot}. Smart homes have also been attacked~\cite{kasp}, even using smart phones~\cite{sivaraman2016smart}. Several medical devices were attacked~\cite{registerinsulin,motherboard}, including medical mannequins~\cite{glisson2015compromising} and pacemakers~\cite{theguardianpacemaker}, resulting in a concern also for the former Vice President of the USA Dick Cheney~\cite{abc}.

IoT devices have also been used to empower ``classical'' attacks. Most notably, in September 2016, the Mirai botnet knocked out a large number of sites including Amazon, Netflix, The New York Times, Reddit, Twitter, Spotify, Playstation, Airbnb, PayPal, and many others using a DDoS attack. The peculiarity of this attack is that the botnet was made up of smart toasters and web-enabled devices~\cite{engadmirai}.

In recent years, security analysts have expanded their focus from cyber-threats to our personal data first to cyber-threats to our devices and then to cyber-threats that may cause direct harm to human beings. For instance,  manufacturers are investing considerable resources and launching bug-hunting projects to prevent attacks to autonomous driving systems (cars, trains, etc.), which are tightly connected with the IoT. Since the consequences of such attacks would be catastrophic, putting millions of lives at risk, the security and trustworthiness of autonomous systems is currently one of the hottest topics in both academia and industry. It has also been considered in the Black Mirror TV series, along with other IoT (in)security scenarios.
For instance, in \emph{Hated in the Nation} (S03E06) artificial substitute bees have been developed to counteract a sudden colony collapse disorder in the UK's bee population... but the robotic bees have been hacked and are being used to kill people.

Some of the dreadful consequences that the IoT could have on our lives are shown in \textit{Arkangel} (S04E02), in which parents can use the Arkangel system to track their children, monitor their health and emotional states, and censor sights they should not see... with devastating effects for parents and children.

In the special \emph{White Christmas}, artificial copies of clients' consciousness, stored into small pods called ``cookies", are used to control smart houses, with the usual Black-Mirror-style disastrous consequences.

The idea of having a copy of ourself as a personal assistant for our houses is a more customized and smarter version of Amazon's \emph{Alexa}, Apple's \emph{Siri}, Google's \emph{Google Assistant} and Microsoft's \emph{Cortana}, intelligent personal assistants that recognize natural voice (in many different languages) without the requirement for keyboard input. These personal assistants support a wide range of user commands, ranging from answering questions to providing real-time information (such as news, weather forecast or traffic information), from making phone calls to compiling to-do lists, from setting alarms to playing music or audiobooks or streaming podcasts, and to acting as a home automation system that controls several other smart devices. See our own~\cite{Gnirut} for a summary of the security risks of such personal assistants, including the possibility of someone hacking one of these assistants to take control of what we usually consider the safest place, our home.\footnote{Similar scenarios have also been considered in several movies. For instance, in \emph{Avengers: Age of Ultron}~\cite{Ultron} an AI called Ultron hacks and apparently destroys Tony Stark's personal assistant J.A.R.V.I.S. taking control of Stark's house.} 

There are two more Black Mirror episodes that are relevant here.
In \emph{The National Anthem} (S01E01), a Princess member of the British royal family has been kidnapped and the kidnapper demands that the country's Prime Minister has sexual intercourse with a pig on live television, otherwise he will kill her. In \emph{Shut up and Dance} (S03E03), 
hackers threaten to leak the blackmail material that they have collected of their victims unless they (the victims) carry out the increasingly despicable tasks that the hackers assign them. These two episodes share the same premise, \emph{blackmail}, which might soon be relevant for the IoT as well. 

\begin{figure}[t]
\begin{center}
\hspace*{-0.55cm}
\includegraphics[scale=0.31]{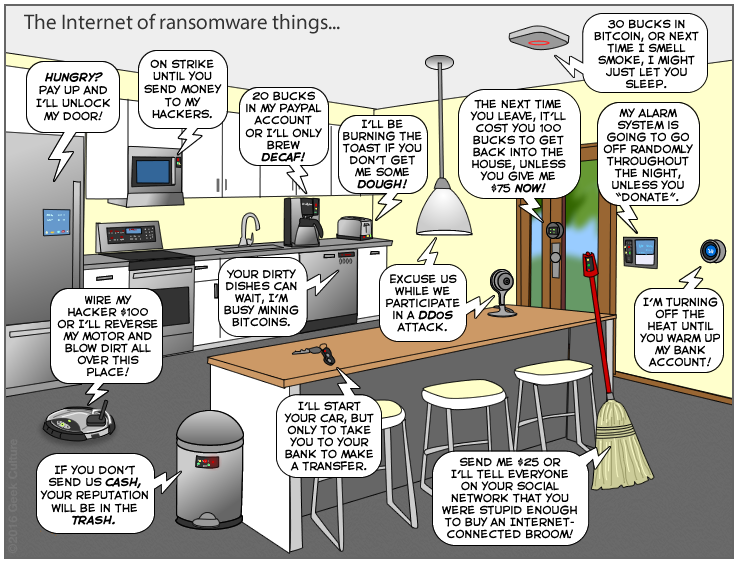}
\end{center}
\vspace*{-0.5cm}
\caption{The Internet of ransomware things... \cite{JoyOfTech}}
\label{fig:IoT-ransomware}
\vspace*{-0.25cm}
\end{figure}

A technological way to blackmail someone is to use \emph{ransomware}. In May 2017, a massive cyber-attack was launched using the ransomware 
WannaCry, which infected more than 230,000 computers in 150 countries~\cite{ehrenfeld2017wannacry}. Once activated, the virus demanded ransom payments in order to unlock the infected system. IoT devices are, and will, of course not be exempt from ransomware attacks and it is just a matter of time before the situations jokingly depicted in the cartoon in Fig.~\ref{fig:IoT-ransomware} become reality.

\section{Smart Humans}
In the rest of this paper, we will discuss a plot-line for a possible future episode of Black Mirror. 
Following the formula that we stated above, it won't be long until prostheses, ECG personal monitors, subcutaneous insulin infusors, glasses, etc. become IoT devices, 
always connected for monitoring, maintenance, charging, tracking but also, why not?, simply because it is cool to control them with a smart phone.
This is the advent of the \emph{Smart Human}, not just a user of the IoT but a Thing in the Internet.\footnote{In~\cite{IoN} (to be presented at ``Re-coding Black Mirror 2019'' as well), we prophesize the forthcoming worldwide deployment of the \emph{Internet of Neurons}, a new Internet paradigm in which humans will be able to connect bi-directionally to the net using only their brain. Smart humans will be a significant step towards the  Internet of Neurons, especially in the first configuration that we consider in~\cite{IoN} in which humans will be implanted a device able to connect their brain to the Internet bi-directionally.} Pros and cons of the IoT will be inherited and to some extent amplified, and new ones will surface, leading to several social, economical, political and ethical implications, including issues related not only to health but also to security and privacy. 

Smart humans are coming and we wonder whether this time we will get it right, designing security in from the start. It might not be too late in the game to do so, but we confess that we are quite skeptical and believe that the security and privacy issues will be massive. So massive, that they have inspired us to conceive of the following Black-Mirror-style scenario:
\emph{What would happen if the Smart in the Human was subject to ransom?}

\section{WannaDie and other security issues}

IoT devices are everywhere. They are around us. They are part of us. We are IoT devices.
How far are we from discovering that hackers are able to hack and control these devices? 
What if someone were able to blackmail us threatening our IoT body parts?
This may be the devastating attack of a new ransomware that we have called \emph{WannaDie}.

\begin{figure}[t]
\hspace*{-1.2cm}
\includegraphics[scale=0.20]{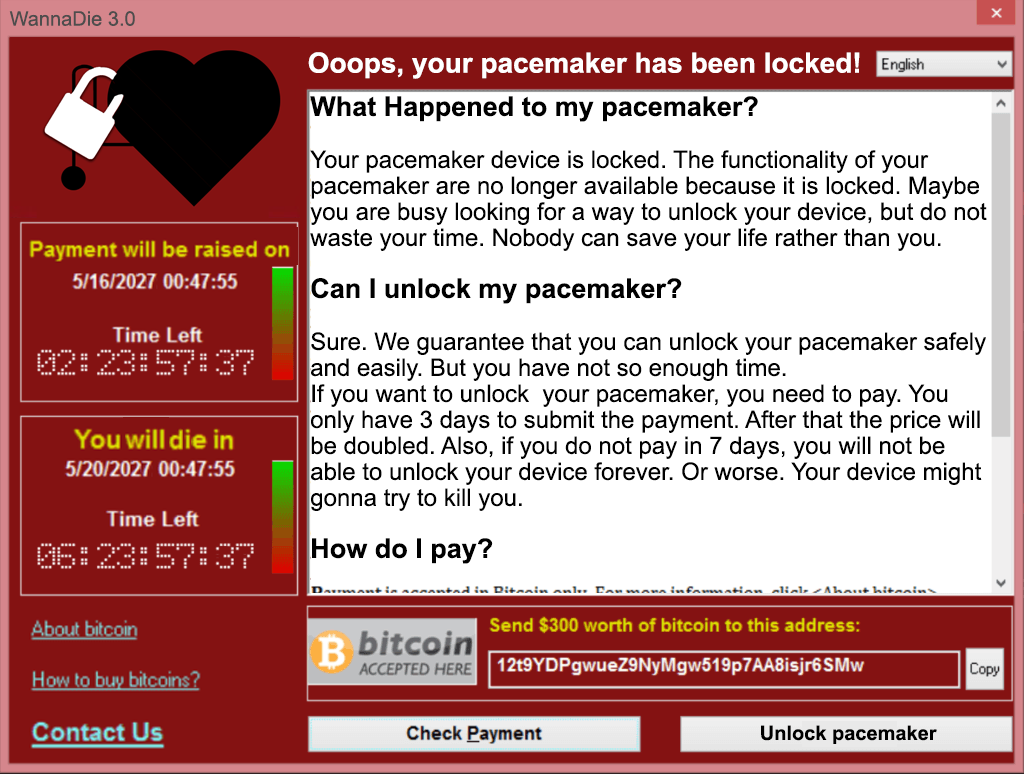}\hskip -4ex
\includegraphics[scale=0.1]{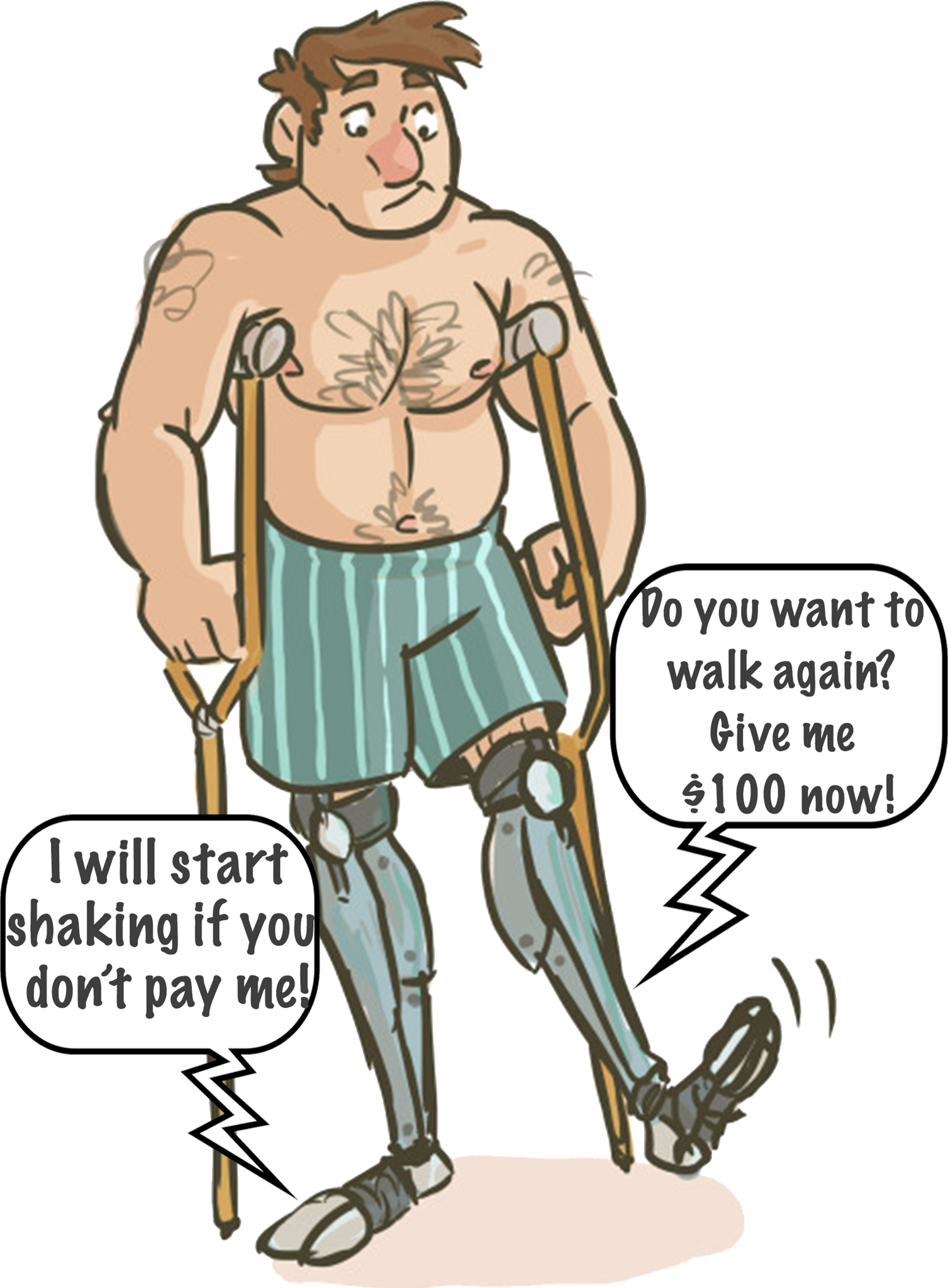}\hspace*{-2cm}
\caption{The WannaDie ransomware at work}
\label{fig:wannadie}
\vspace*{-0.5cm}
\end{figure}

WannaDie's victims might receive a message like the one on the left of Fig.~\ref{fig:wannadie} announcing that their fully-IoT-connected pacemaker has been be locked, or it could be their legs as shown on the right of that figure.\footnote{While we don't really care whether WannaCry's authors will mind that we have taken their message and adapted it to WannaDie, we hope that Nitrozac \& Snaggy won't mind that we have taken their comic as inspiration for our drawing shown on the right of the figure (which we modified from the original~\cite{man} to include the balloons of the legs speaking).}

As we observed above, for Smart Humans to be secure, it will be necessary to build in security from the start. Devices should be developed by applying \emph{security-by-design} principles~\cite{owasp} to avoid inheriting typical IoT weaknesses such as diversity of devices, standards and environments)~\cite{7167238,hossain2015towards}.
As security-by-design is most often wishful thinking, let us briefly discuss some contingency security procedures that could be put in place. 

WannaCry spread widely and rapidly thanks to the slowness of systems around the world in applying security patches: the exploit used (a Windows' Server Message Block protocol flaw) was already patched about two months earlier. 
For the security of Smart Humans, we will need to create an infrastructure that allows software \emph{updates} to be applied rapidly and thereby prevent an infection. 
In parallel to that, it would be wise to allow humans to intervene once  the infection has started. This could be achieved by means of an ``off/on'' button to bypass normal functioning, but of course this would require some form of strong authentication (e.g., biometric, via the voice of the human) to access the button in ``root mode''. Note, however, that resetting or rebooting a pacemaker might be dangerous.

Hackers could also try to ransom money by capitalizing on the reputation of a previous attack, as it happened recently with WannaCry itself~\cite{wannback}. While most of us who received such ``phishing'' emails ignored them safely, in the case of WannaDie people might not want to take that chance worrying that their life might be at risk.

So far, humans got sick with diseases, not malware. However, removing the separation between devices and humans, as is in fact the case when the human becomes a Thing, opens scenarios for new kinds of cyber-health threats. This requires revising and broadening the notions of security and privacy. \emph{Location privacy}, for instance, will be difficult to achieve if a human has an implanted device that is always connected. Humans might also want to keep hidden the fact that they have an implant, a property that we could call \textit{smart privacy}.

All these security and privacy issues could be exploited to enrich the plot of a new Black Mirror episode. But also the practical problems of Smart Humanity will contribute to making it interesting, and frightful.
Solutions to problems like connectivity and power will be crucial. Will the devices use WiFi to be connected? Yes, as long as a WiFi connection is available. However, devices could also take advantage of 5G connectivity (or 6G, by then), but this won't be without creating new risks for security and health issues due to the close exposure to electromagnetic fields. To keep their functionalities and the connectivity always on, such devices will need to be recharged often, if not constantly, due to the batteries' life and size. A solution could be to use a wireless power transmission as in~\cite{chabalko2017quasistatic}, but would it still be feasible if the device is inside our body? 

Finally, one could envision also a number of psychological issues.\footnote{As we discuss in~\cite{SchroedingersMan} (to be presented at ``Re-coding Black Mirror 2019'' as well), there are also philosophical and metaphysical issues, e.g., related to the identity paradox of the \emph{Ship of Theseus}. For instance, the Japanese manga and animated series \emph{Ghost in the Shell} cyclically returns to the paradox of a ``human'' in which people have their organic body parts replaced by artificial parts, sometimes going so far as to have their entire body replaced with a prosthetic one, leaving the brain as the only remaining original part. Is the result still the same human being?}
Nowadays humans sometimes reject implants. This is mainly due to their body rejecting the ``alien component'' but in some cases due to the psychological challenge of accepting the implant and the change it requires both in the personal and in the social sphere; this could be exacerbated by an IoT implant up to the point that some people might prefer not to have it at all, even when they need it. There could even be \emph{anti-smarter} movements.

It is time to take stock.
Like every momentous change, the advent of Smart Humans will go through different phases. It will initially encounter resistance and people will be reluctant to get on board. A slow acceptance phase will follow, in which people will begin to embrace the idea of Smart Humans, thus leading to trust and diffusion of the technology for the geeks, the curious and the wealthy. It is unclear to us if the technology will ever be available to everybody, and if so, if it will be free. History teaches us that this is typically not the case, so we also expect that a new area of social studies will be born to investigate the social implications of Smart Humans interacting with each other but especially with those who are still ``smart-less'', which will cause discrimination and division. 

Smart humans will not just provide the premise of another discomforting and scary Black Mirror episode, but it will also force us to reconsider the term ``humanity'', trying to identify what may be the line that separates humanity from its possible technological evolution(s).

\end{document}